\documentclass[12pt,preprint]{aastex}

\newcommand{\kev}{\mbox{$\rm\,keV$}}
\newcommand{\cm}{\mbox{$\rm\,cm$}}

\slugcomment{Submitted to ApJ on 12/Jun/2006}

\shorttitle{Spectral Signatures of Photon Bubbles}
\shortauthors{Finke \& B\"ottcher}

\begin{document}


\title{X-ray Spectral Signatures of the Photon Bubble Model for 
Ultraluminous X-ray Sources}


\author{Justin D. Finke and Markus B\"ottcher}
\affil{Astrophysical Institute, Department of Physics and Astronomy, \\
Ohio University, Athens, OH 45701}
\email{finke@helios.phy.ohiou.edu, mboett@helios.phy.ohiou.edu}



\begin{abstract}
The nature of ultraluminous X-ray sources in nearby galaxies is one of
the major open questions in modern X-ray astrophysics. One possible 
explanation for these objects is an inhomogeneous, radiation dominated 
accretion disk around a $\sim 10 M_{\odot}$ black hole --- the so-called
``photon bubble'' model. While previous studies of this model have 
focused primarily on its radiation-hydrodynamics aspects, in this
paper, we provide an analysis of its X-ray spectral (continuum and possible
edge and line) characteristics. Compton reflection between high and 
low density regions in the disk may provide the key to distinguishing 
this model from others, such as accretion onto an intermediate mass black
hole. We couple a Monte Carlo/Fokker-Planck radiation transport 
code with the \-XSTAR code for reflection to simulate the 
photon spectra produced in a photon bubble model for ULXs.  
We find that reflection components tend to be very weak and 
in most cases not observable, and make predictions for the shape 
of the high-energy Comptonizing spectra.  In many cases the 
Comptonization dominates the spectra even down to $\sim$ a few 
keV.  In one simulation, a $\sim 9 \kev$ feature was found, 
which may be considered a signature of photon bubbles in ULXs; 
furthermore, we make predictions of high energy power-laws 
which may be observed by future instruments.

\end{abstract}


%% subject header.  Objects should be in the appropriate "individual"
%% headers (e.g. quasars: individual, stars: individual, etc.) with the
%% additional provision that the total number of headers, including each
%% individual object, not exceed six.  The \objectname{} macro, and its
%% alias \object{}, is used to mark each object.  The macro takes the object
%% name as its primary argument.  This name will appear in the paper
%% and serve as the link's anchor in the electronic edition if the name
%% is recognized by the data centers.  The macro also takes an optional
%% argument in parentheses in cases where the data center identification
%% differs from what is to be printed in the paper.

\keywords{accretion, accretion disks --- methods:  numerical --- 
radiative transfer --- X-rays:  binaries}

%individual(\objectname{NGC 6397},
%\object{NGC 6624}, \objectname[M 15]{NGC 7078},
%\object[Cl 1938-341]{Terzan 8})}


\section{Introduction}
\label{intro}

Approximately 150 off-nuclear Ultraluminous X-ray Sources (ULXs) in
nearby galaxies have been discovered with luminosities greater than 
$10^{39}$ erg s$^{-1}$, 
exceeding the Eddington Luminosity for
a $\sim 10 M_{\odot}$ black hole \citep[e.g.,][]{fabbiano88,cm99,cp02,lm05}.  
While some can be identified as supernova remnants, background Active Galactic
Nuclei, or faint foreground stars \citep[e.g.,][]{g06}, most seem to be the 
result of accretion from a high-mass star onto a compact object. Short-term
variability of some of these ULXs indicates that they are not simply 
unresolved superpositions of several lower-luminosity sources 
\citep[e.g.,][]{matsumoto01,fabbiano03}. Thus, the high 
luminosity implies accretion onto black holes with masses 
$50 M_{\odot} < M < 10^4 M_{\odot}$,
\citep[intermediate mass black holes; IMBHs;][]{cm99,metal00,hui05,metal06} 
or super--Eddington accretion.  Super--Eddington accretion can be achieved
in two different ways:  an inhomogeneous accretion-disk structure in which 
the photon flux is spatially separated from the bulk of the matter influx, 
or strongly anisotropic radiation. 

Anisotropic emission \citep{king01} may originate, e.g., from jet sources
associated with accretion onto solar-mass black holes (microquasars)
in which the jet is oriented at a small angle with respect to our line
of sight --- the so-called ``microblazar'' model \citep{gak02,kfm02}.   
The latter model is now considered unlikely based on recent observations 
of X-ray ionization of optical nebulae associated with some ULXs 
\citep{pm03, g06}.

A promising mechanism for sustaining 
super-Eddington accretion in a radiation-dom\-in\-at\-ed, magnetized accretion 
flow, is provided by the so-called photon bubble instability 
\citep{a92,g98,begelman01,begelman02,begelman06}. 
In this model, radiation can escape 
along low-density regions (LDRs) aligned with predominantly vertical
magnetic field lines.  The accretion flow is concentrated in thin,
optically thick high-density regions (HDRs), where the magnetic pressure
of field lines oriented predominantly within the disk dominates over 
the gas pressure, thus confining the gas and preventing the radiation
pressure from disrupting the accretion flow. In such a configuration,
the total disk luminosity can exceed the Eddington luminosity by a
factor approximately equal to the ratio of the magnetic field pressure
to the gas pressure \citep{begelman01,begelman02} in the HDRs. 

Given the variety of different promising candidate models for the nature
of ULXs, the obvious question is: Can one distinguish between these models 
with ULXs' X-ray spectra? If so, what are the characteristic spectral features
of the individual models? From an accreting 
IMBH it is generally believed that a multi-color 
disk blackbody (MCDBB) spectrum with an inner temperature of $kT_{in} \sim$ 
0.1 --- 0.3~keV (that is, the disk temperature at the innermost stable 
circular orbit) and possibly a high energy component 
from Compton up-scattering 
in a tenuous, hot corona would be a realistic phenomenological description
of the spectrum. However, a recent, more detailed analysis of non-LTE
accretion flows around IMBHs \citep{hui05} has indicated that the effects
of black-hole rotation and Compton scattering within the disk may very well
lead to much higher apparent disk temperatures, up to $kT \sim 1$~keV,
in addition to deviations from conventional MCDBB spectra at both soft
and hard X-ray energies due to metal opacity effects. 

The continuum spectrum from the photon bubble model might be dominated 
by the MCDBB spectrum emanating from the
HDRs.
However, this spectrum might be modified during the 
radiation transport in the photon bubble cavities. Furthermore, the almost 
free-streaming radiation of the photon bubbles could be repeatedly 
Compton-reflected off the surfaces of the HDRs, 
potentially leading to strong fluorescence lines and/or radiative 
recombination edges, in addition to Compton reflection features from 
hard X-ray emission impinging upon the disk from radiation sources
external to the disk \citep{btb04,bty05}. Such features may be 
observable, assuming it is not overwhelmed by other radiation sources, 
in particular the blackbody from the HDRs. 

Distinguishing scenarios by spectral modeling is currently very difficult 
for all but the highest quality ULX data sets, and spectral fitting to 
these ULX spectra gives contradictory results. For example, \citet{fk05}
recently performed a detailed spectral and timing analysis of archival 
{\it XMM-Newton} data on 28 ULXs that had sufficient photon 
statistics to allow for meaningful fitting with models more 
complicated than a simple power-law.  They found that their continuum 
spectra fell into three general categories: (1) Optically thin 
bremsstrahlung-dominated, quasi-thermal spectra with temperatures 
of $kT \sim 0.6$ -- 0.8~keV, characteristic for X-ray emission from 
young supernova remnants; (2) MCDBB spectra of temperatures $kT \sim 
0.1$ -- 0.4~keV, plus occasionally a hard X-ray power-law, as possible 
in the case of accretion onto IMBHs; (3) MCDBB spectra at temperatures 
$kT \sim 1$~keV, plus a power-law component dominating at lower energies. 
Based on these results, 
\cite{fk05} suggest that ULXs may, in fact, not be a homogeneous 
class of objects. This is 
in accord with recent results of \cite{metal06}
that realistic stellar evolution and population synthesis calculations
suggest that the expected rate of captures of massive stars by IMBHs
may not be sufficient to produce the total observed number of ULXs. 
\citet{wetal06} classified ULXs into low/hard and high/soft 
states based on their X-ray luminosities and spectra, assuming the
ULXs were IMBHs accreting in states similar to galactic X-ray binaries.
They found that the high/soft ULXs were grouped around
one of two blackbody temperatures:  one grouped around $~1$ keV and one
grouped around $~0.1$ keV, further indication that ULXs are
not a homogeneous group. The long term monitoring of some ULXs such
as NGC 5204 X-1 shows that its X-ray spectrum hardens as its flux
increases, opposite to what has been observed in Galactic X-ray
binaries \citep{robetal06}.

\citet{srw06} found that 6 out of 13 of the most-observed ULXs were 
fit approximately equally as well with a cool MCDBB and hard power-law 
as with a soft power-law and warm MCDBB.  Even more surprisingly, 
they found that most spectra (10 out of 13) were fit best by a cool, 
$kT \sim 0.2$~keV blackbody and a warmer $kT_{in} \sim 2$~keV MCDBB, 
and 11 out of 13 were fit by a cool MCDBB and an optically thick 
($\tau \sim 10$) corona. These spectra could also be explained by 
a hot, inner (optically thick) plasma sphere and a cool outer disk 
\citep{am06}, in which 
case the cool blackbody temperature could not be directly related 
to the black hole mass by virtue of the relation between the inner 
disk temperature and the black hole mass. 

In summary, current diagnostics for the distinction between different
models for ULXs indicate that ULXs may in fact not be a homogeneous
set of physical objects, different models may apply to different
sources, and in many cases, observational results are inconclusive. 
Consequently, more detailed model predictions of various ULX models
might be helpful in identifying additional diagnostics which may be
used to confirm or rule out such models. In this paper, we investigate
detailed spectral features resulting from the photon bubble model,
both in the continuum and possible emission line and radiative
recombination edge features resulting from the radiative feedback
between the HDRs and LDRs. An Fe K line has 
been observed in M82 X-1 \citep{sm03,am06}; absorption edges have 
been observed in the 0.1 -- 1~keV range in M101 ULX-1 \citep{ketal04}; 
and possible edges at $\sim 0.7$ keV and $\sim 8$ keV, and an
emission feature at $\sim 6$~keV has been seen in Holmberg IX
X-1 \citep{dgr06}.  In the ULX M51 X-26 emission features
have been seen at 1.8~keV, 3.24~keV, 4.03~keV, and 6.65~keV with 
{\it Chandra} \citep{tw04} and at 6.4~keV with {\it XMM--Newton}
\citep{detal05}. This indicates that the detection of emission
lines and/or radiative recombination edges is currently feasible,
at least for bright ULXs, and predictions of expected line features
from various models will thus be useful as an additional model
diagnostic.  

Studies of X-ray reflection in inhomogeneous accretion disks
have been done by
\citet{rfb02} and \citet{fetal02}, who found that multiple 
Compton reflections of emission from a corona above
an irregular or corrugated accretion disk in Seyfert galaxies
lead to much stronger reflection features than single reflections.  
\citet{meretal06} considered
Compton reflection in a radiation--dominated inhomogeneous
accretion disk.  Their disk was a low density Comptonizing plasma with
high density clumps off of which radiation was reflected.  Their 
simulations produced lines which may be visible in narrow line Seyfert 
1 galaxies, although detection may be difficult due to 
interstellar  absorption. 
So far, investigations of photon bubbles in ULXs have focused
on radiation--hydrodynamics \citep[see, e.g., ][]{rb03, tetal05} 
leaving the expected spectral features essentially unexplored.
\citet{btb04} and \citet{bty05} studyed X-ray reflection off accretion
disks in the photon bubble model for Seyfert galaxies, but using
an unspecified external irradiating hard X-ray source, such as a 
tenuous, hot corona above the accretion disk surface, while we are 
focusing on the self-consistent local radiation feedback between the 
HDRs and LDRs within the inhomogeneous accretion flow. 
Furthermore, those authors did not explore parameters
appropriate for ULXs. 

In this paper we make predictions of the X-ray
spectra from the photon bubble model of ULXs with a 
Monte Carlo/Fokker-Planck code coupled with the XSTAR program for 
X-ray reprocessing in the HDRs (reflection).
We will be primarily interested in predictions
of spectral features in addition to a warm MCDBB and soft power-law
component, although this is not a study of any individual object.  
In addition we make predictions of a power-law component 
that extends above 10 keV which may not be detectable by 
{\em Chandra} or {\em XMM-Newton} but may be by {\em Suzaku} or 
future telescopes.  

\section{Model Setup}
\label{setup}

The simulation parameters were based on solutions found in 
\citet{begelman01} and \citet{begelman02}.  
We summarize these solutions in \S\ \ref{disk},
and describe the simulation technique in \S\ \ref{simulation}.  

\subsection{Disk Distribution}
\label{disk}

The physical picture emerging from numerical simulations of the
photon bubble instability \citep{rb03,tetal05} points towards
a propagating pattern of slab-like shock trains (the 
HDRs) slanted with respect to the plane of the accretion disk.
A solution to the disk structure has been derived by
\citet{begelman02}, who found the relevant disk parameters to depend on:
the accretion rate, $\dot{m} = \dot{M}\kappa c/4\pi GM_{BH}$, 
$\beta$, the ratio of the HDR's gas pressure 
to the magnetic pressure, and $\alpha$, the Shakura-Sunyaev viscosity
parameter.  Here 
$\kappa$ is the opacity (assumed to be dominated by 
Thomson scattering; $\kappa=$
0.4 cm$^2$ g$^{-1}$), $c$ is the
speed of light, $G$ is the gravitational constant, and $M_{BH}$ is 
the mass of the black hole; We used his solution to set up our simulations.
Throughout this work we assumed $\beta = 0.1$ and 
$m = M_{BH} / M_{\odot} = 10$ (resulting in an Eddington luminosity of 
$1.3 \cdot 10^{39}$ erg s$^{-1}$).

The Eddington enhancement factor, the ratio of the disk's flux
to the Eddington flux, is:
\begin{equation}
\label{ratio}
l \equiv \frac{F}{F_{Edd}} = \frac{n_{HDR}}{n_{avg}} 
= \frac{3}{2} \frac{\dot{m}D}{\delta r}
\end{equation}
where $n_{avg} = \sqrt{n_{HDR}n_{LDR}}$ is the geometric 
average of the LDR and HDR densities; $r$ is the distance from the
black hole in terms of gravitational radii ($R_g = GM/c^2$); 
$\delta$ is the ratio of the height to the radius ($\delta = h/R$); 
and $D =  1 - (6/r)^{1/2}$.  \citet{begelman02} derives the following
equation for $\delta$:
$$
\delta \sim  \max \Biggl[  0.3  \left( \frac{\beta}{0.1} \right) ^{-4/13} 
\left( \frac{\alpha}{0.01} \right) ^{-5/13} 
(\dot{m}D)^{5/13} m^{-1/13} r^{-5/26}, 
$$
\begin{equation}
\label{delta}
%\begin{eqnarray}
 0.2   \left( \frac{\beta}{0.1} \right)^{4/21} 
\left( \frac{\alpha}{0.01} \right)^{-5/21}
(\dot{m}D)^{5/21} m^{-1/21} r^{1/14} \Biggr]
%\end{eqnarray}
\end{equation}
Note that in Eqn. \ref{delta} 
we corrected a typographical error in Eqn. 13 of
\citet{begelman02}.  
Based on \citet{begelman02}'s expression for the average density, and
equation \ref{ratio}, the following equations were derived for
the densities in the HDR and LDR:
\begin{equation}
\label{HDR}
n_{HDR} = 6.0\cdot10^{18}(\dot{m}D)^2(m\alpha)^{-1}\delta^{-4}r^{-5/2}\ 
          \cm^{-3}
\end{equation}
\begin{equation}
\label{LDR}
n_{LDR} = 3.6\cdot10^{18}(m\alpha)^{-1}\delta^{-2}r^{-1/2}\ 
          \cm^{-3}.
\end{equation}
Although these densities refer to the disk midplane, the \citet{ss73} 
solution for a radiation-dominated accretion disk shows that the 
density will be essentially constant in the vertical direction.  
Thus, we consider these densities to be the density at the 
photosphere.
The temperature for the HDRs was found by assuming that all the 
gravitational potential energy
is released in the HDRs, and thus, the flux emitted at the 
photoplane is set equal to the blackbody
flux.  This gives the \citet{ss73} expression:
\begin{equation}
\label{tempHDR}
kT_{HDR} = 5.09\ m^{-1/4} (\dot{m}D)^{1/4} r^{-3/4}\ \kev.
\end{equation}

The ``wavelength'' of the plane-parallel shocks---i.e., this 
distance between HDRs---is given by 
\begin{equation}
\lambda = \beta\alpha\delta R.  
\end{equation}

For the proton temperature in the LDRs, the virial temperature 
was used:
\begin{equation}
\label{kTp}
kT_{lp} = \frac{GM_{BH}m_p}{R}
\end{equation}

where $m_p$ is the proton's mass.  

The electron temperature in the LDR is expected to be significantly
lower than the proton temperature due to Compton cooling by radiation
from the HDRs.  The LDR's electron temperature
was determined by implicitly solving a
Fokker-Planck equation with the MC/FP code; 
see \S\ \ref{simulation} below.  The magnetic field in the 
LDR was determined by assuming it to be in equipartition with the
electrons.

The Eddington ratio is dependent on radius (equation \ref{ratio}), 
so for a given disk, outside of a certain radius ($r_{inhom}$) 
$l<1$, and the 
disk becomes a homogeneous, Shakura-Sunyaev disk.  Outside of this
radius, the disk was represented by MCDBB with
$kT \propto r^{-3/4}$.
 
\subsection{Simulation Description}
\label{simulation}

Equations \ref{delta} through \ref{kTp} 
were used to determine
the simulation parameters at various radii.  To find the total spectrum
of the disk, simulations were run
at evenly-spaced radii and 
the results were averaged, weighted by disk area.  
At each radius, we simulated one individual LDR 
sandwiched by two HDRs in a plane-parallel geometry 
(see Fig. \ref{geometry}).  Within the plane-parallel geometry the 
LDR is divided into 40 zones, 4 radial and 10 vertical.
It seems reasonable to assume, that, since 
$\lambda$ is much smaller than the radiation 
pressure scale height \citep{begelman02}, that the fraction of photons which 
escape (given by the ratio of the escape area to the region's 
total surface area), will be $\sim 1$\%.  Larger escape fractions 
could lead to smaller ionization parameters and greater reflection components,
however, this was not explored in this work.  
Multiple Compton reflections are certainly possible with such
a small escape fraction, however for nearly all of our cases, 
the reflected component was completely drowned out by the 
blackbody; see below.  As a result, further iterations did not 
produce significantly different spectra.  

The radiation transfer and self-consistent
balance between electron heating and cooling within the LDR are 
simulated with the two-dimensional
Monte Carlo/Fokker-Planck
(MC/FP) code described in \citet{bjl03} and \citet{bl01}.  
This code 
uses the Monte Carlo method of \citet{pss83} for Compton scattering
and the implicit Fokker-Planck method of \citet{nm98} for the
evolution
of the electron distribution in a two-temperature plasma with a 
given proton temperature.  In each zone, 
the Fokker-Planck equation is implicitly and independently solved.  
The Fokker-Planck technique takes into
account heating/cooling by Coulomb/M\o ller interactions, Compton scattering,
synchrotron/cyclotron processes.  

Initially, the emission emanating from the HDRs was represented by 
blackbody spectra inserted at the upper and lower boundaries.  
Photons were also produced by synchrotron/cyclotron processes
in the LDRs.  Photons were subjected to Compton scattering and 
reflection off the HDRs.
Escaping photons at the outer boundaries were added to an event 
file for later spectral extraction; this involves
placing each escaping photon in a particular energy bin.  

The proton temperature in the LDR was calculated from equation
\ref{kTp}, and the electron temperature
was calculated numerically within 
the MC/FP simulation \citep[see][]{nm98,bjl03,bl01,fb05}, which was run 
until the electron temperature reached a stable equilibrium.
The MC/FP
simulation was run to extract the photon flux and spectrum 
incident on the boundaries between the HDRs and the LDR. To 
calculate the expected spectral features from fluorescence
line emission, radiative recombination, and Compton reflection, 
the impinging spectrum was used as an input into the latest 
version of XSTAR \citep{kb01}. Solar abundances of the most 
profuse astrophysical elements (H, He, C, Ca, N, O, Ne, Mg, 
Si, S, Ar, Ca and Fe) were assumed in the HDR, based on \citet{gns96}.
XSTAR was run in constant pressure mode; the pressure was calculated 
from the specified density and the ideal gas law. 
The main parameter in determining the reflection spectrum's shape 
and intensity is the ionization parameter:
\begin{equation}
\label{ionparam}
\xi = \frac{4 \pi F}{ n_{HDR} }.
\end{equation}
Note that XSTAR uses the flux calculated between 1 and 1000 Ry 
(13.6 eV to 13.6 keV).
In order to 
circumvent XSTAR's limitation to densities $n_e \lesssim 
1 \cdot 10^{17}$cm$^{-3}$, the impinging flux was re-scaled
to keep the ionization $\xi$ at the 
value corresponding to the physical situation. 
The inverse flux
scaling was applied to the XSTAR output spectrum.  Keeping 
$\xi$ constant insures that the scaling of a flux dominated by
recombination features, $F \propto n_e$ is properly recovered.  
Testing of XSTAR with various densities but the same $\xi$ 
seems shows 
that this is appropriate for the energy range of interest.   
The resulting 
reflection spectrum was then added to the intrinsic blackbody 
from the HDR as boundary sources in a second run of the MC/FP 
code for the final evaluation of the emanating X-ray spectrum.
For an example of a simulation, with spectra at different
radii, see Fig. \ref{specradii}.

\section{Results}
\label{results}

The simulation parameters and fit results 
can be seen in Table \ref{parameters}.
The important disk parameters and are plotted as a function of radius
in Figs. \ref{paramradius1} to \ref{paramradius3}.  

The transition between the expressions for $\delta$ (Eqn. \ref{delta})
can be readily seen in panel (a) of Fig. \ref{paramradius1},
with the first expression being used at lower radii.  At higher
$\alpha$, the transition occurs at lower radii, or not at all.
Note that several simulations violate the thin-disk condition 
($\delta < 1$), especially the high $\dot{m}$ and/or low
$\alpha$ ones.  These solutions should thus be viewed with caution.

The ionization parameter decreases with radius, as 
is observed in our simulations, seen in panel (b) of Figs. 
\ref{paramradius1} to \ref{paramradius3}.  
At no point does $\xi$ drop below $\sim 2000$ erg cm s$^{-1}$, 
so the reflection component will be small ($\la 12\%\ $
 of the total) for all cases.  

The optical depth, $\tau$, is proportional to $\lambda n_{LDR}$, and is 
essentially always increasing with $r$.  Thus the spectra have 
more luminosity from Comptonization at larger radii.  It 
is also larger for smaller accretion rates due to its $\delta$
dependence 
($\tau \propto \lambda n_{LDR} \propto \delta n_{LDR} \propto \delta^{-1}$)
by virtue of $\delta$  increasing with increasing $\dot{m}$.
Thus we expect a greater Comptonization component for 
smaller $\dot{m}$.
We can see that for our simulations, this is in fact the case 
(Fig. \ref{spectot}).  We can also see in 
Table \ref{parameters} that the fractional power emitted through
Comptonization decreases with increasing $\dot{m}$, as expected.  
This also corresponds to the decrease in $\Gamma$, the 
photon index ($EL_E \propto E^{-\Gamma+2}$) with increasing 
$\dot{m}$.  One can also see in Figs. 
\ref{paramradius1}---\ref{paramradius3} that as $\alpha$ increases, 
$\tau$ increases.  Again, this is because 
$\tau \propto \delta^{-1}$, and as $\alpha$ increases, $\delta$ 
decreases.  

The ratio $l$ peaks at $r\sim 12$,
as can be seen in panel (d) of
Figs. \ref{paramradius1}-- \ref{paramradius3}.  This is 
close to where the peak flux is expected in a typical 
Shakura-Sunyaev accretion disk.  Also note that $l$ 
increases with increasing $\alpha$, as one would expect from 
eqn. \ref{ratio} ($l\propto \delta^{-1}$).

For each simulation, the total reflection component from XSTAR 
was summed up 
for all radii.  The soft X-ray spectra were fit with a MCDBB.
The MCDBB and reflection components were subtracted from the total
spectra revealing the Comptonization components.  This component was 
fit with a power-law and exponential cutoff 
($EL_E \propto E^{-\Gamma+2}\ e^{-E/E_{cutoff}}$) above 25 keV.  
The results of this 
decomposition and fits are summarized in Table \ref{parameters};
the individual fit components for the $\alpha=0.01$ simulations
can be seen in Fig. \ref{reflect}.  
Note that these fit components are the sum of the
simulations at all radii.  
The only simulation that has a significant reflection 
feature visible in its spectrum is Sim. 2, with a blended 
Fe XXV/XXVI feature at $\sim 9\ \kev$ with an equivalent width of 
$\sim 640$ eV, and an edge at $\sim 0.87\ \kev$ (O K).  This 
simulation also has the highest fraction of its emission 
from the reflection component.  One can see in Fig. \ref{specradii} 
that the reflection component is strongest at $r=120$---280.  

In Fig. \ref{specradii}, one can see
that the Comptonization component gets stronger at 
larger radii, as one would expect with the increasing 
optical depth.  
The overall spectra are generally dominated by the 
high-energy Comptonization component, especially for higher
$\dot{m}$.  
The Comptonization component in many cases extends 
far down in photon energy into the $\sim$ few keV range 
and often dominates the total bolometric luminosity, 
even when the total spectrum appears to be dominated 
by a thermal soft X-ray component.

This component contributes to the spectra even 
at low energies, down to $\sim 1 \kev$, as seen in 
Fig. \ref{reflect}.  In Fig. \ref{specradii}, one can see
that the Comptonization component gets stronger at 
larger radii, as one would expect with the increasing 
optical depth.  This explains the significant Comptonization 
component to the lower energy part of the spectrum.  
Eventually, though, one gets to the radius where photon bubbles 
can no longer be sustained ($r_{inhom}$), and there is 
no longer a Comptonization component, as seen in $r>520$ in 
Fig. \ref{specradii}.  Results for other simulations are similar 
to Fig. \ref{specradii}.  As $\alpha$ increases, the 
$r_{inhom}$ increases, as can be seen in Figs. 
\ref{paramradius1}---\ref{paramradius3} and 
Table \ref{parameters}.  

Spectral pivoting can be seen in the all of the 
simulations in Fig. \ref{spectot}.  
This is due to the approximately
same amount of luminosity in the Compton components of the different 
simulations.  The energy dissipated in the LDRs is related to the 
proton temperature, which in our simulations does not depend on 
accretion rate.  This is not an unrealistic assumption, as one 
would not expect the protons to radiate significantly.  
However, when the LDRs become more optically thick 
the energy dissipation becomes more efficient due to multiple 
Compton scatterings.  

\section{Discussion}
\label{discussion}

Our simulations explore 
0.5--10 keV luminosities up to $\sim 8 \cdot 10^{39}$ erg s$^{-1}$
which does not include some of the brightest ULXs.  For our 
setup, luminosities much greater than this will violate 
the thin disk condition, except possibly for the highest value 
of $\alpha$.  It does not seem likely that photon bubbles could 
support much greater luminosities.  

Our results are significantly different from previous, similar 
works.  \citet{meretal06} perform similar calculations for a 
low density accretion region with imbedded optically thick 
clumps and considered reflection between these clumps.  
They used ionization parameters comparable to ours 
($\sim 3000$ erg cm s$^{-1}$, among others) and 
achieved much stronger reflection components with no 
$\sim 9 \kev$ feature.  
This is a consequence of the substantially lower temperature 
of the soft blackbody (ultraviolet) photons used as an 
input into the Comptonization scheme in their model:  their 
Comptonization component extends down to $\la 0.1 \kev$ 
energies (in contrast to $\sim$ a few keV in our case).  
For comparable $\xi$, calculated from the 1---1000 Ry flux 
the number of photons above the Fe K edge is thus much
smaller than in our case.  Therefore, heavy metals 
are almost fully ionized in most of our simulations, leading 
to weak or absent line features and a relatively weak Compton
reflection component.
\citet{btb04} 
explored reflections from a hot corona off of an inhomogeneous 
disk and found the spectrum can differ significantly from 
reflection off a homogeneous disk.  They also had stronger 
reflection features than our results, due to 
lower ionization parameters, and no $\sim 9\ \kev$ 
feature.  

\citet{fk05} and \citet{srw06} fit ULXs' X-ray spectra 
with various models; when fit with a MCDBB or MCDBB and 
low-energy blackbody, their fits had inner disk temperatures 
similar to ours.  However, we find that our model is unable to 
explain a soft-excess.  We note that no $\sim 9 \kev$ 
feature has been detected in the spectra of ULXs, although 
features at other energies have been detected 
\citep{sm03,am06,ketal04,dgr06}, although these 
features could originate from a wind rather than 
reflection features in the disk.  ULX observations 
above $\sim 10 \kev$ by, {\em Suzaku} or the next generation
of hard X-ray imaging instruments might be able to 
detect the hard power-laws we predict, although
we realize such observations would be difficult.  
{\em Suzaku} observations of two ULXs in NGC 1313 
did not detect any component above 10 keV above the background 
\citep{mizuno07}, although their MCDBB + power-law fits do seem 
to agree with our spectra, and the variability of these sources is 
much less than observed in Galactic black hole candidates.  

Unfortunately, our simulations do not reproduce 
a soft excess as has been observed in many ULXs.  It is 
possible that the soft excess originates from 
Compton downscattering of radiation by wind \citep[e.g.,][]{begelman01}, 
and so a 
lack of its production in our simulations does not 
disprove the photon bubble model for ULXs.  

If the excess is explained, our model could explain 
ULXs well fit with a $\sim 1$ keV blackbody.  
Higher energy observations ($\ga10$ keV) could 
determine this; if they found a stable hard X-ray power-law 
$\Gamma \sim$ 2---2.4 for a long period of time, this would be 
evidence for the photon bubble model in ULXs, due to the 
fact that a lower mass compact object would have less 
variability \citep{kalogera04}.  
A $\sim 9 \kev$ feature could also be considered a signature 
of photon bubbles in ULXs; it has not been found in any other 
simulation of accreting black holes.  

We note that our spectra are similar to the very high state 
of X-ray binaries such as GX 339-4 \citep{belloni06} and 
GRO J1655-40 \citep{saito06}, which have similar 
photon indices as found in our simulations.  
For smaller accretion rates, the photon bubble
model may be a viable model to explain the very high state 
of X-ray binaries.

Since this is an exploratory study, we have made a number of 
simplifying assumptions.  
We have assumed a plane-parallel geometry;
zones could be corrugated or vary in random ways, 
which could lead to multiple reflections, increasing the reflection
component \citep{fetal02}.  
We have assumed constant densities across
LDRs.  This may alter the Comptonizing region of the spectra, 
but probably would not lead to significant differences.
We have completely neglected several items.
Taking into account 
General Relativistic effects could gravitationally broaden 
the spectrum.  However, most of the emission originates at 
$r\sim12$, too far out to be greatly affected by GR.  
We have assumed $n_{HDR}$ and $n_{LDR}$ to be constant in the 
vertical direction, neglecting color corrections; 
this may be validated by the more detailed 
vertical solution of \citet{begelman06}, who found 
that the corrections should be minor.    
Time variability is beyond the scope of this study 
and is poorly understood, and thus would involve 
poorly constrained parameters.
It is also possible 
that the spectra could be modified by further Comptonization in 
a corona above the disk, which we have also neglected.  This 
would make the spectra even harder than they already are, 
possibly too hard to match observations.  However, recent 
magnetohydrodynamic simulations have shown difficulties in 
creating coronae \citep{hirose06} and the photon bubble 
model may be an alternative to the standard disk-corona 
geometry.  Future simulations could take advantage of 
more detailed analytic solutions \citep[e.g.,][]{begelman06} 
and hydrodynamic simulations \citep[e.g.,][]{tetal05} 
and could include reprocessing in a corona and/or 
a disk wind.

\acknowledgments{
This work was partially supported by NASA through {\it XMM-Newton}
GO grant no. NNG04GI50G and INTEGRAL theory grant NNG05GK59G, 
as well as by an allocation of computing time from the 
Ohio Supercomputer Center (OSC) via grant PHS0256-1.  
Simulations were run on the OSC Pentium 4 Cluster in Columbus, Ohio.
We thank the referee for many helpful suggestions which have 
improved this paper.

\clearpage
\begin{deluxetable}{cccccccccccc}
\rotate
\tabletypesize{\scriptsize}
\tablecaption{Simulation and fit parameters.  $\dot{m}$ is the 
accretion rate, $\alpha$ is the Shakura-Sunyaev viscosity parameter, 
$kT_{in}$ is the inner disk temperature, and $r_{inhom}$ is the
radius above which photon bubbles cannot exist ($l<1$).  
Results of our simulations are:
total luminosity over the Eddington luminosity is given 
($L_{tot} / L_{Edd}$); the 0.2---10 keV luminosity over the 
Eddington luminosity ($L_X / L{Edd}$);
the fitted photon index, $\Gamma$; 
the high energy exponential
cutoff; and the fraction of the 
total luminosity in the MCDBB component($f_{BB}$), in the Comptonization 
component ($f_{Compt}$), and the reflection component ($f_{refl}$).  
Note that these do not all add up to unity due to rounding.
}
\tablewidth{0pt}
\tablehead{ 
\colhead{ Sim. No. } &
\colhead{ $\dot{m}$ } & 
\colhead{ $\alpha$ } &
\colhead{ $kT_{in}$ [keV] } &
\colhead{ $r_{inhom}$ } &
\colhead{ $L_{tot} / L_{Edd}$ } & 
\colhead{ $L_X / L{Edd}$ } &
\colhead{ $\Gamma$ } &
\colhead{ $E_{cutoff}$ [keV] } &
\colhead{ $f_{BB}$ } &
\colhead{ $f_{Compt}$ } &
\colhead{ $f_{refl}$ }
} 
\startdata
1  & 200 & 0.01 & 1.4 &  280 & 2.8  & 1.3 &  2.18 & 460 & 0.22 & 0.73 & 0.05 \\
2  & 400 & 0.01 & 1.6 &  472 & 4.4  & 2.5 &  2.42 & 440 & 0.26 & 0.63 & 0.11 \\
3  & 900 & 0.01 & 2.0 &  860 & 6.9  & 4.7 &  2.42 & 290 & 0.56 & 0.40 & 0.05 \\
4  & 200 & 0.1  & 1.4 &  482 & 5.0  & 1.7 &  1.96 & 280 & 0.17 & 0.82 & 0.01 \\
5  & 400 & 0.1  & 1.6 &  804 & 6.5  & 2.9 &  2.12 & 220 & 0.27 & 0.74 & 0.01 \\
6  & 900 & 0.1  & 2.0 & 1456 & 9.4  & 5.3 &  2.24 & 140 & 0.39 & 0.59 & 0.02 \\
7  & 200 & 0.5  & 1.4 &  698 & 5.8  & 1.8 &  1.75 & 190 & 0.14 & 0.85 & 0.01 \\
8  & 400 & 0.5  & 1.6 & 1164 & 8.5  & 3.1 &  1.78 & 160 & 0.17 & 0.79 & 0.05 \\
9  & 900 & 0.5  & 2.0 & 2100 & 11.9 & 5.6 &  1.95 & 160 & 0.31 & 0.68 & 0.02 \\
\enddata
\label{parameters}
\end{deluxetable}
\clearpage

%% the file, uncomment the \plotone, \plottwo, and
%% If you need a layout that cannot be achieved with \plotone or
%% \plottwo, you can invoke the graphicx package directly with the
%% \includegraphics command or use \plotfiddle. For more information,
%% please see the tutorial on "Using Electronic Art with AASTeX" in the
%% documentation section at the AASTeX Web site,
%% http://www.journals.uchicago.edu/AAS/AASTeX.
%%
%% The examples below also include sample markup for submission of
%% supplemental electronic materials. As always, be sure to check
%% the instructions to authors for the journal you are submitting to
%% for specific submissions guidelines as they vary from

\begin{figure}
\epsscale{1.0}
\plotone{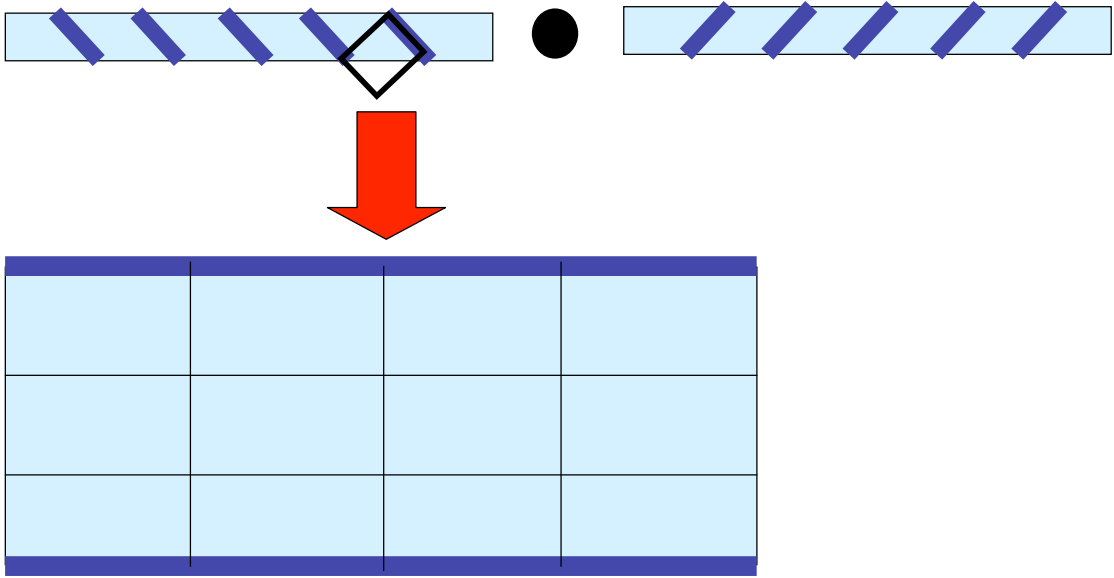}%{f1.eps}
%\plotone{geometry2.eps}
\caption{The simulation geometry.  Each simulation is of one part of
the disk.  The LDR is divided into 40 zones.  Afterwords, simulations 
of different parts of the disk are averaged, weighted by area.
See the electronic edition of the Journal for a color version of this figure.
\label{geometry}}
\end{figure}

\clearpage

\begin{figure}
\epsscale{1.0}
\plotone{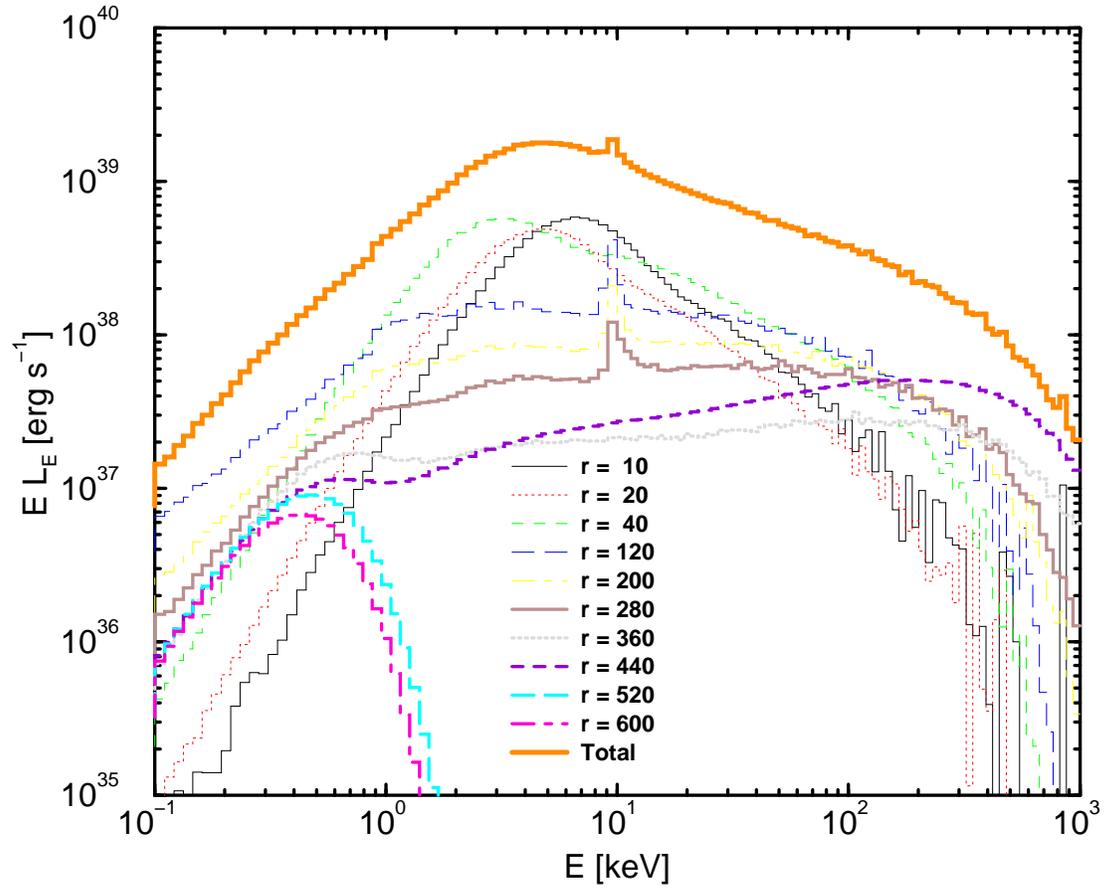}%{f2.eps}
%\plotone{specradii.eps}
\caption{Spectra at various radii and the total spectrum for 
Simulation 2.  See the electronic edition of the Journal for a color 
version of this figure.  }
\label{specradii}
\end{figure}

\clearpage

\begin{figure}
\epsscale{1.0}
\plotone{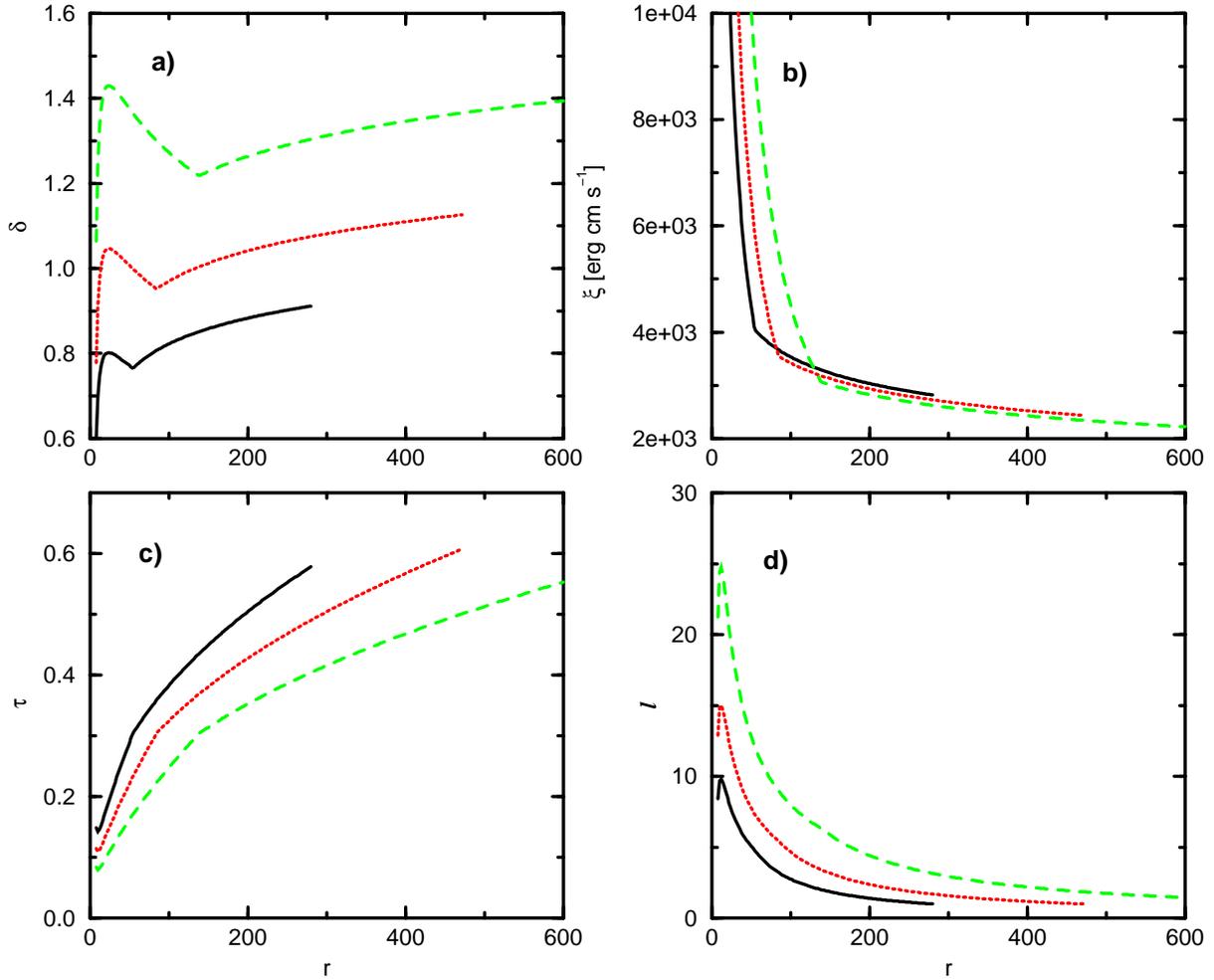}%{f3.eps}
%\plotone{paramradius1.eps}
\caption{The parameters $\delta$ (a), $\xi$ (b), $\tau$ (c), 
and $l$ (d) as a 
function of radius, for $\dot{m}=200$ (solid black), $\dot{m}=400$ 
(dotted red), 
and $\dot{m}=900$ (dashed green).  These simulations have $\alpha=0.01$.  
See the electronic edition of the Journal for a color version of this figure.}
\label{paramradius1}
\end{figure}

\clearpage

\begin{figure}
\epsscale{1.0}
\plotone{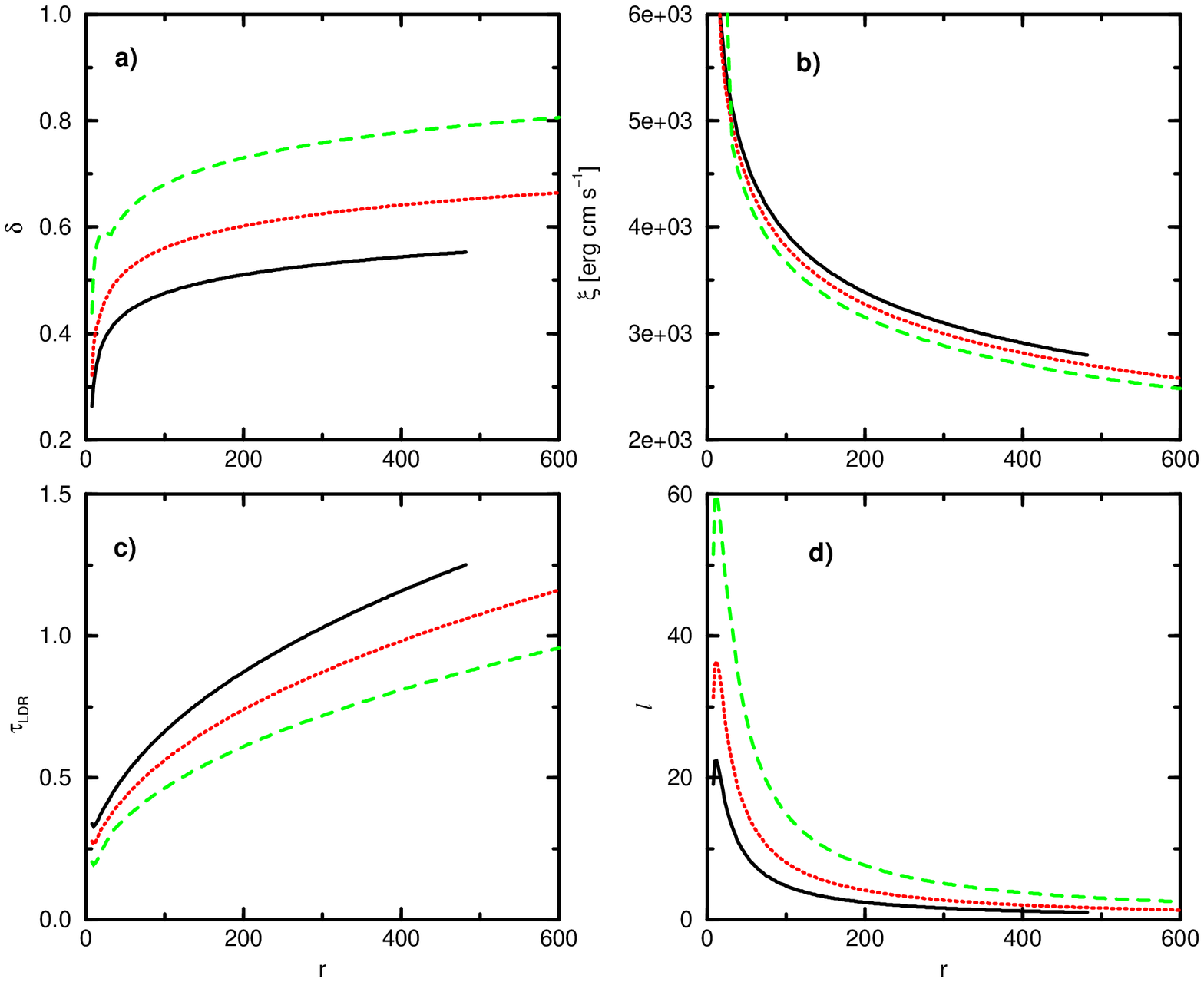}%{f4.eps}
%\plotone{paramradius2.eps}
\caption{Same as Fig. \ref{paramradius1} except for $\alpha=0.1$.
See the electronic edition of the Journal for a color version of this figure.}
\label{paramradius2}
\end{figure}

\clearpage

\begin{figure}
\epsscale{1.0}
\plotone{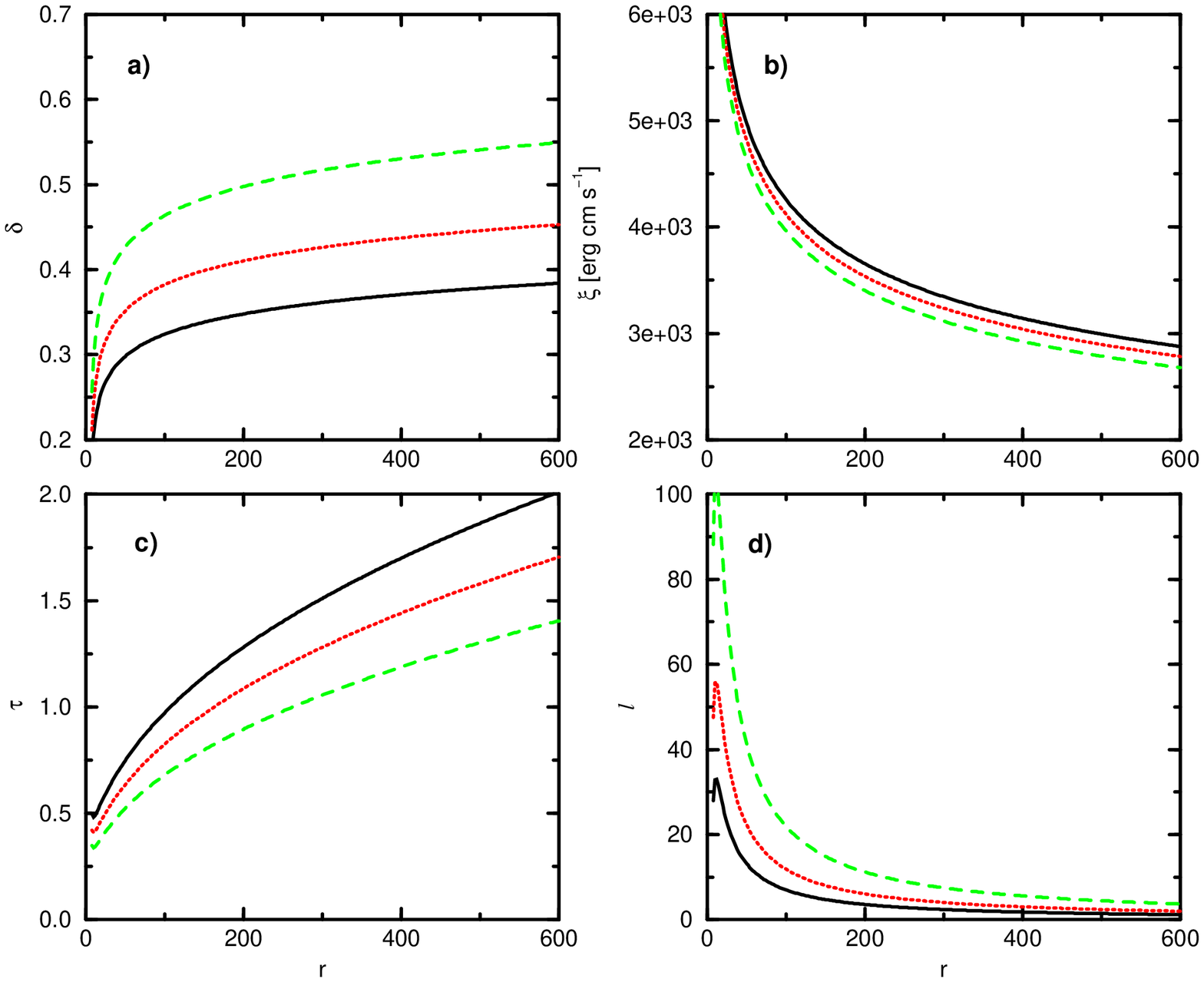}%{f5.eps
%\plotone{paramradius3.eps}
\caption{Same as Fig. \ref{paramradius1} except for $\alpha=0.5$.
See the electronic edition of the Journal for a color version of this figure.}
\label{paramradius3}
\end{figure}

\clearpage

\begin{figure}
\epsscale{1.0}
\plotone{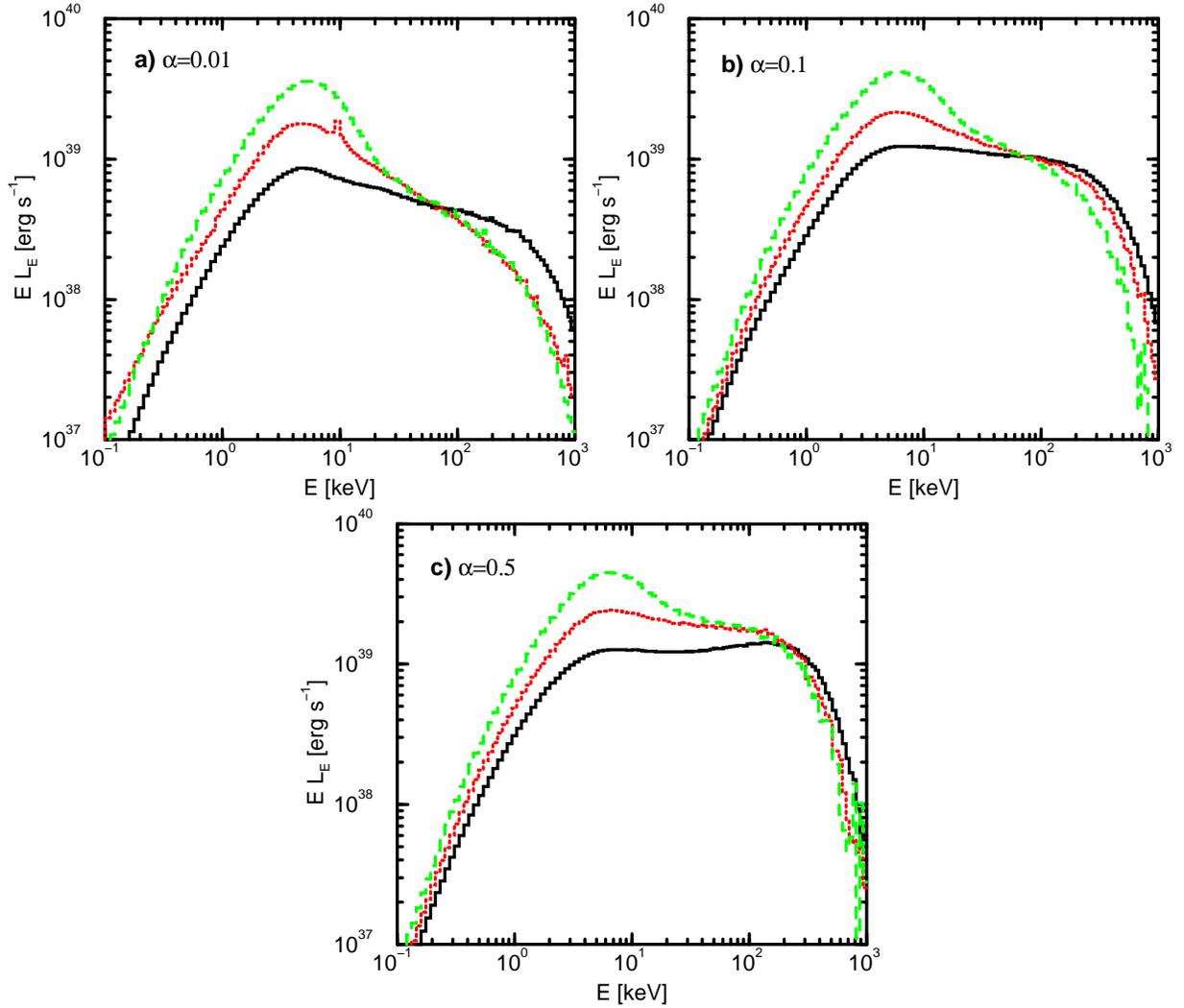}%{f6.eps}
%\plotone{spectot.eps}
\caption{Total spectra for (a) $\alpha=0.01$, (b) $\alpha=0.1$, and
(c) $\alpha=0.5$.  For all graphs we plot $\dot{m}=200$ (solid black), 
$\dot{m}=400$ (dotted red), and $\dot{m}=900$ (dashed green).  
See the electronic edition of the Journal for a color version of this 
figure.  }
\label{spectot}
\end{figure}

\clearpage

\begin{figure}
\epsscale{1.0}
\plotone{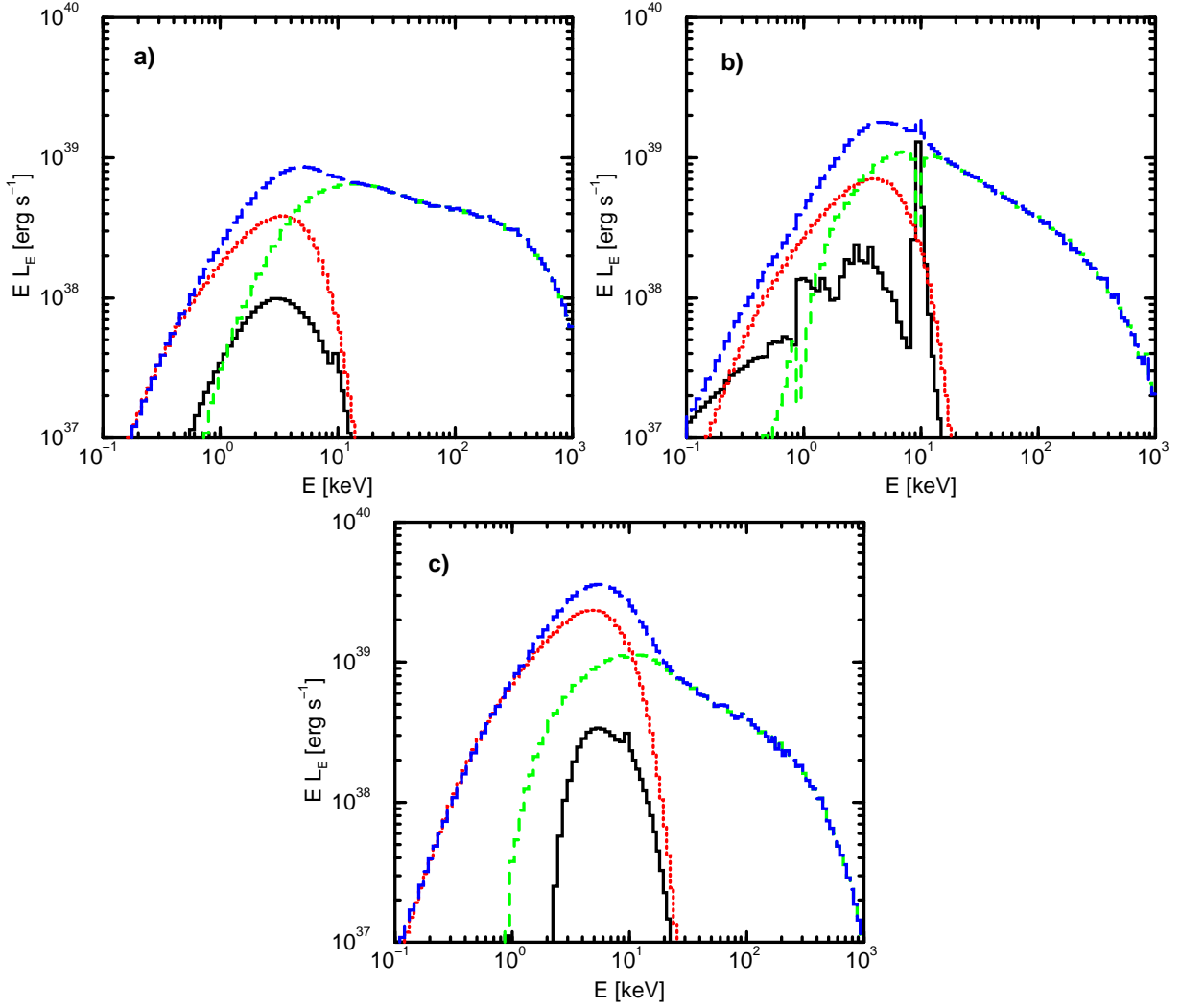}%{f7.eps}
%\plotone{reflect.eps}
\caption{The spectra broken into components for $\alpha=0.01$ and 
(a) $\dot{m}=200$, (b) $\dot{m}=400$, and (c) $\dot{m}=900$.  
The solid black line is the reflected component, the dotted red line is the 
MCDBB, the dashed green line is the Comptonization 
component, and the long dashed blue line is the total spectrum. 
See the electronic edition of the Journal for a color version of this 
figure.  }
\label{reflect}
\end{figure}

\clearpage

\end{document}